# AIS-INMACA: A Novel Integrated MACA Based Clonal Classifier for Protein Coding and Promoter Region Prediction

**Pokkuluri Kiran Sree[1], Inampudi Ramesh Babu[2] and SSSN Usha Devi N[3]**

[1]Research Scholar, Department of CSE, JNTU Hyderabad, India
[2]Professor, Department of CSE, ANU, Guntur, India
[3]Assistant Professor, Department of CSE, University College of Engineering, JNTUK, India

**\*Corresponding author:** Pokkuluri Kiran Sree, Department of CSE, JNTU Hyderabad, India; Email: profkiransree@gmail.com

**Abstract**

Most of the problems in bioinformatics are now the challenges in computing. This paper aims at building a classifier based on Multiple Attractor Cellular Automata (MACA) which uses fuzzy logic. It is strengthened with an artificial Immune System Technique (AIS), Clonal algorithm for identifying a protein coding and promoter region in a given DNA sequence. The proposed classifier is named as AIS-INMACA introduces a novel concept to combine CA with artificial immune system to produce a better classifier which can address major problems in bioinformatics. This will be the first integrated algorithm which can predict both promoter and protein coding regions. To obtain good fitness rules the basic concept of Clonal selection algorithm was used. The proposed classifier can handle DNA sequences of lengths 54,108,162,252,354. This classifier gives the exact boundaries of both protein and promoter regions with an average accuracy of 89.6%. This classifier was tested with 97,000 data components which were taken from Fickett & Toung , MPromDb, and other sequences from a renowned medical university. This proposed classifier can handle huge data sets and can find protein and promoter regions even in mixed and overlapped DNA sequences. This work also aims at identifying the logicality between the major problems in bioinformatics and tries to obtaining a common frame work for addressing major problems in bioinformatics like protein structure prediction, RNA structure prediction, predicting the splicing pattern of any primary transcript and analysis of information content in DNA, RNA, protein sequences and structure. This work will attract more researchers towards application of CA as a potential pattern classifier to many important problems in bioinformatics.

## Introduction

### Cellular automata

Pattern classification encompasses development of a model which will be trained to solve a given problem with the help of some examples; each of them will be characterized by a number of features. The development of such a system is characterized as pattern classification. We use a class of Cellular Automata (CA)[1,2] to develop the proposed classifier. Cellular automata consist of a grid of cells with a finite number of states. Cellular Automata (CA) is a computing model which provides a good platform for performing complex computations with the available local information.

CA is defined a four tuple <G, Z, N, F>

Where

G -> Grid (Set of cells)
Z -> Set of possible cell states
N -> Set which describe cells neighborhoods
F -> Transition Function (Rules of automata)

As Cellular Automata consists of a number of cells structured in the form of a grid. The transitions between the cells may depend on its own state and the states of its neighboring cells. The equation one sates that if ith cell have to make a transition, it has to depend on own state, left neighbor and right neighbor also.

$$q_i(t+1) = f(q_{i-1}(t), q_i(t), q_{i+1}(t)) \text{ ---- Equation 1}$$

## Problems in Bioinformatics

Bioinformatics can be characterized as a collection of statistical, mathematical and computational methods for dissecting biological sequences like DNA, RNA and amino acid. It deals with the design and development of computer based technology that supports biological processing. Bioinformatics tools are aimed at performing lot of functions like data integration, collection, analysis, mining, management, simulation, visualization and statistics.

The central dogma shown in the Figure 1 of molecular biology was initially articulated by Francis Crick [3] in 1958. It deals with point by point transfer of important sequential data. It states that data can't be exchanged back from protein to either protein or nucleic acid. So once data gets into protein, it can't stream again to nucleic acid. This dogma is a framework for comprehension about the exchange of sequence data between sequential data carrying biopolymers in living organisms. There are 3 significant classes of such biopolymers ie DNA, RNA and protein.

There are 9 possible immediate exchanges of data that can happen between these. This dogma classifies these nine exchanges into three transfers (Normal, Special, and Never Happens). The normal flow of biological information is DNA is copied to DNA (DNA replication), DNA information is copied into RNA (Transcription) and protein are synthesized by using the protein coding region exits in DNA/RNA (Translation).

## Protein Coding Region Identification

DNA is an important component of a cell and genes will be found in specific portion of DNA which will contain the information as explicit sequences of bases (A, G, C, T).These explicit sequences of nucleotides will have instructions to build the proteins. But the region which will have the instructions which is called as protein coding regions occupies very less space in a DNA sequence. The identification of protein coding regions plays a vital role in understanding the gens. We can extract lot of information like what is the disease causing gene, whether it is inherited from father or mother, how one cell is going to control another cell.

## Promoter Region Identification

DNA is a very important component in a cell, which is located in the nucleus. DNA contains lot of information. For DNA sequence to transcript and form RNA which copies the required information, we need a promoter. So promoter plays a vital role in DNA transcription. It is defined as "the sequence in the region of the upstream of the transcriptional start site (TSS)". If we identify the promoter region we can extract information regarding gene expression patterns, cell specificity and development. Some of the genetic diseases which are associated with variations in promoters are asthma, beta thalassemia and rubinstein-taybi syndrome.

## Literature Survey

Eric E. Snyder et al. [4,5] has developed a PC program, Geneparser, which distinguishes and verifies the protein coding genes in genomic DNA arrangements.

This program scores all subintervals in a grouping for substance facts characteristic of introns and exons, and for destinations that recognize their limits. Jonathan H. Badger[6] et al. has proposed a protein coding region identification tool named CRITICA which uses comparative analysis. In the comparative segment of the investigation, regions of DNA are straightened with identified successions from the DNA databases; if the interpretation of the arranged groupings has more stupendous amino acid than needed for the watched rate nucleotide character; this is translated as proof for coding.

David J. States[7] proposed a PC project called BLASTX was formerly indicated to be viable in distinguishing and allotting putative capacity to likely protein coding districts by identifying huge likeness between a theoretically interpreted nucleotide query arrangement and parts of a protein grouping database.

Steven Salzberg, et al. [8] has used a decision tree algorithm for locating protein coding region. Genes in eukaryotic DNA spread hundreds or many base sets, while the locales of the aforementioned genes that code for proteins may possess just a little rate of the succession. Eric E.Snyder, et al. [4] has used dynamic programming with neural networks to address the protein coding region problem. Dynamic Programming (DP) is connected to the issue of exactly distinguishing inside exons and introns in genomic DNA arrangements. Suprakash Datta, et al. [9] used a DFT based gene prediction for addressing this problem. Authors provided theoretical concept of three periodicity property observed in protein coding regions in genomic DNA. They proposed new criteria for classification based on traditional frequency approaches of coding regions. Jesus P. Mena-Chalco, et al. [10] has used Modified Gabor-Wavelet Transform for addressing this issue. In this connection, numerous coding DNA model-free systems dependent upon the event of particular examples of nucleotides at coding areas have been proposed. Regardless, these techniques have not been totally suitable because of their reliance on an observationally predefined window length needed [11] for a nearby dissection of a DNA locale. Many authors[10,12-15,16] have applied CA in bioinformatics.

Rakesh Mishra, et al. [17] has worked on search and use of promoter region. Look for a promoter component by RNA polymerase from the to a great degree extensive DNA base arrangement is thought to be the slowest and rate-confirming for the regulation of interpretation process. Few immediate investigations we portrayed here which have attempted to accompany the robotic suggestions of this promoter look[18]. Christoph Dieterich, et al. [19] made an extensive study on promoter region. The robotized annotation of promoter districts joins data of two sorts. To begin with, it recognizes cross-species preservation inside upstream districts of orthologous genes. Pair wise too a various arrangement examinations are processed. Vetriselvi Rangnanan, et al. [20] made a dissection of different anticipated structural lands of promoter [19] districts in prokaryotic and in addition eukaryotic genomes had prior shown that they have a few normal characteristics, for example, lower steadiness, higher curve and less bendability, when contrasted and their neighbouring areas [20]. Jih-Wei Hung[21] has developed an effective forecast calculation that can expand the recognition (power =1 - false negative) of promoter. Authors introduce two strategies that utilize the machine force to ascertain all conceivable examples which are the conceivable characteristics of promoters. Some of the other authors [22-26] has worked on promoter identification and succeeded to some extent.

## AIS-INMACA(Artificial Immune System-Integrated Multiple Attractor Cellular Automata)

Multiple Attractor Cellular Automata which is used in this report is a special class of fuzzy cellular automata which was introduced thirty years ago. It uses fuzzy logic [27,28] to handle real value attributes. The development process / implementation of Multiple Attractor Cellular Automata is administered by AIS technique, a Clonal Algorithm with the underlying theory of survival of the fittest gene.

Artificial Immune System is a novel computational intelligence technique with features like distributed computing, fault /error tolerance, dynamic learning, adaption to the frame work, self monitoring, non uniformity and several features of natural immune systems. AIS take its motivation from the standard immune system of the body to propose novel computing tools for addressing many problems in wide domain areas. Some features of AIS which can be mapped with bioinformatics framework are chosen and used in the thesis to strengthen the proposed CA classifier.

This paper introduces the integration of AIS with CA which is first for its kind to produce a better classifier which can address major problems in bioinformatics. This proposed classifier named artificial immune system based multiple attractor cellular automata classifier (AIS-INMACA) uses the basic frame work of Cellular Automata (CA) and features of AIS like self monitoring and non uniformity which is potential, versatile and robust. This is the basic motivation of the entire research.

The objectives of AIS based evolution of MACA [13-14, 28] is
1. To improve the conception of the ways CA performs calculations.
2. To Figure out how CA may be advanced to perform a particular computational job.
3. To follow how advancement makes complex global behavior into locally interconnected cells on a grid.
4. To extract innate classification potential in CA and use Clonal algorithm for producing better rules with fitness.

## Design of AIS-INMACA

The proposed AIS based CA classifier uses fuzzy logic can address major problems inbioinformatics like protein coding region identification and promoter region prediction .Even though some scientists have proposed different algorithms, all of these are specific to the problem. None of them have worked towards proposing a common classifier whose frame work can be useful for addressing many problems in bioinformatics.

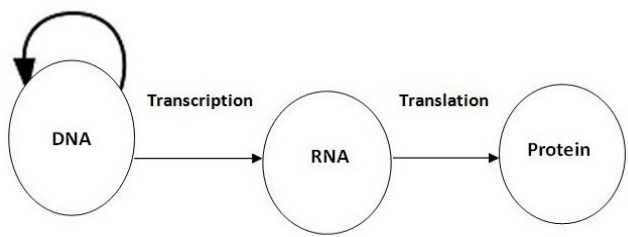

**Figure 1:** Central Dogma

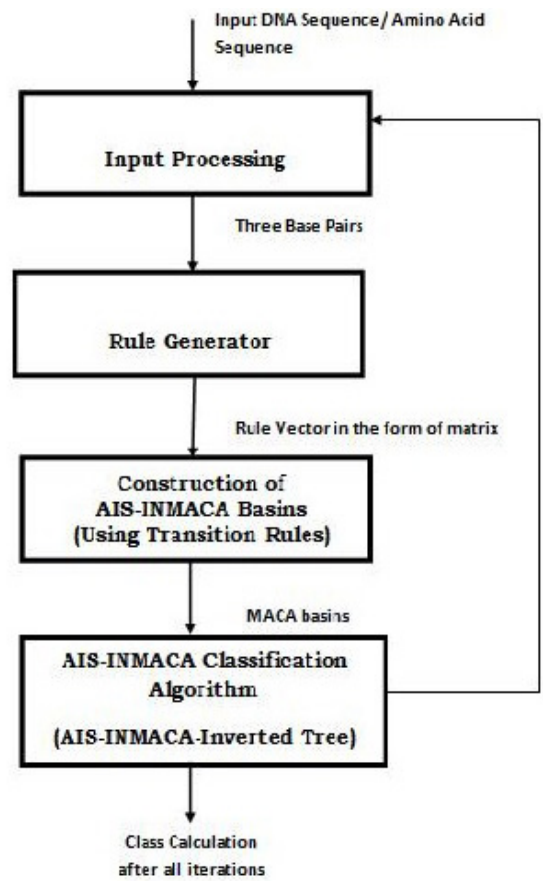

**Figure 2:** Design of AIS-INMACA

| Neighbourhood | 111 | 110 | 101 | 100 | 011 | 010 | 001 | 000 | Rule |
|---|---|---|---|---|---|---|---|---|---|
| Next State (Rule 51) | 0 | 0 | 1 | 1 | 0 | 0 | 1 | 1 | 51 |
| Next State (Rule 254) | 1 | 1 | 1 | 1 | 1 | 1 | 1 | 0 | 254 |

**Table 1:** Transition Function

| SNO | Rule Number | General Representation |
|---|---|---|
| 1 | 1 | $\overline{q_{i-1} + q_i + q_{i+1}}$ |
| 2 | 3 | $\overline{q_{i-1} + q_i}$ |
| 3 | 17 | $\overline{q_i + q_{i+1}}$ |
| 4 | 5 | $\overline{q_{i-1} + q_{i+1}}$ |
| 5 | 51 | $\overline{q_i}$ |
| 6 | 15 | $\overline{q_{i-1}}$ |
| 7 | 85 | $\overline{q_{i+1}}$ |
| 8 | 255 | 1 |

**Table 2:** Complemented Rules

| SNO | Rule Number | General Representation |
|---|---|---|
| 1 | 254 | $q_{i-1} + q_i + q_{i+1}$ |
| 2 | 252 | $q_{i-1} + q_i$ |
| 3 | 238 | $q_i + q_{i+1}$ |
| 4 | 250 | $q_{i-1} + q_{i+1}$ |
| 5 | 204 | $q_i$ |
| 6 | 240 | $q_{i-1}$ |
| 7 | 170 | $q_{i+1}$ |
| 8 | 0 | 0 |

**Table: 3:** Non complemented Rules

The general design of AIS-INMACA is indicated in the Figure. 2 Input to AIS-INMACA algorithm and its variations will be DNA sequence and Amino Acid sequences. Input processing unit will process sequences three at a time as three neighborhood cellular automata is considered for processing DNA sequences. The rule generator will transform the complemented (Table 2) and non complemented rules (Table 3) in the form of matrix, so that we can apply the rules to the corresponding sequence positions very easily. AIS-INMACA basins are calculated as per the instructions of proposed algorithm and an inverter tree named as AIS multiple attractor cellular automata is formed which can predict the class of the input after all iterations.

Algorithm 3.1 is used for creating of AIS-INMACA tree .This tree will dissipate the DNA sequence into respective leaves of the tree. If the sequence is falling into two or more class labels, the algorithm wills recursively partition [29] in such a way that all the sequences will fit into one of the leaves. Every leaf will have a class .Algorithm 2 will be used for getting the class as well as the required transition function. The best fitness rules with a score more than .5 is considered. Algorithm 3.3 [7] uses, algorithm 3.1, 3.2 for predicting the protein and promoter coding regions.

### Rules of AIS-INMACA

The decimal equivalent of the next state function, as defined as the rule number of the CA cell introduced by Wolfram [2], is. In a 2-state 3-neighborhood CA, there are 256 distinct next state functions, among 256 rules, rule 51 and rule 254 are represented in the following equations.

$$\text{Rule 51}: q_i(t+1) = \overline{q_i(t)} \quad (1)$$
$$\text{Rule 254}: q_i(t+1) = q_{i-1}(t) + q_i(t) + q_{i+1}(t) \quad (2)$$

The transition function Table 1 was shown for equations 1,2. The tables 2, 3 show the complemented and non complemented rules.

### Algorithm 3. 1:

Input: Training Set S= {S1, S2,………….., Sx) with P classes

Output: AIS-INMACA tree

Partition(S, P)

1. Generate a AIS-INMACA with x attractor basins (Two Neighborhood CA)
2. Distribute training set in x attractor basins(Nodes)
3. Evaluate the patterns distributed in each attractor basin.
4. If all the patters say S' which are covered by the attractor basin belong to only one class, then label the attractor basin ( Leaf Node) as the class
5. If S' of an attractor basin belong to more than one class partition (S',P')
6. Stop

### Algorithm 3.2 (Partial CLA Algorithm)

Input: Training set S = {S1, S2, • • • , SK},   Maximum Generation (Gmax).

Output: Dependency matrix T, F, and class information.
begin

Step 1: Generate 200 new chromosomes for IP.
Step 2: Initialize generation counter GC=zero; Present Population (PP) ← IP.
Step 3: Compute fitness Ft for each chromosome of PP according to Equation
Step 4: Store T, F, and corresponding class information for which the fitness value Ft>0.5
Step 5: If number of chromosomes with fitness more than 0.5 are 50  then go to 12.
Step 6: Rank chromosomes in order of fitness.
Step 6a: Clone the chromosome
Step 7: Increment generation counter (GC).
Step 8: If GC > Gmax then go to Step 11.
Step 9: Form NP by selection, cloning and mutation.
Step 10: PP← NP; go to Step 3.
Step 11: Store T, F, and corresponding class information for which the fitness value is maximum.
Step 11a: Output class, T,F

### Algorithm 3.3:

1. Uses the AIS-INMACA Tree construction Algorithm 3.1
2. Uses the AIS-INMACA Evolution Algorithm 3. 2
3. Trace the corresponding attractor
4. Travel back form attractor to the starting node
5. Identify the start codon
6. Identify the stop codon
7. Report the boundaries of protein coding region.
8.  From fist codon of the sequence to start codon Search for TAATAA.
9. Report the promoter boundary located at upstream

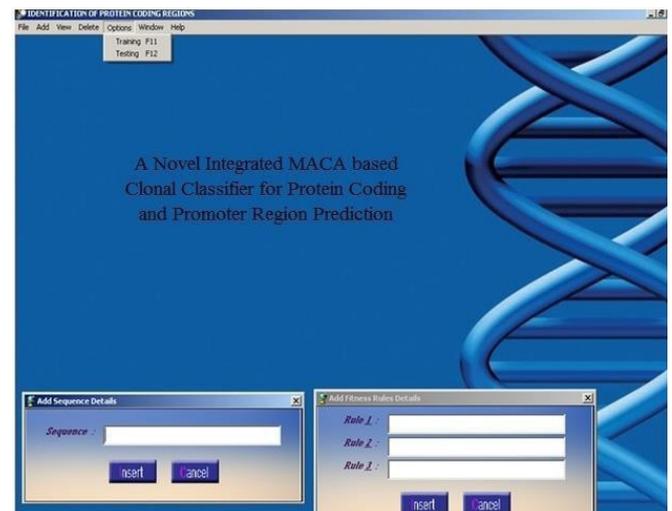

**Figure 3:** Interface

### Experimental Results & Discussions

Experiments were conducted by using Fickett and Toung data [30] for predicting the protein coding regions. All the 21 measures reported in [30] were considered for developing the classifier. Promoters are tested and trained with MpromDb data sets[31]. Figure 3 shows the interface developed. Figure. 4 is the training interface with rules, sequence and real values. Figure 5 shows the testing interface. Table. 4 show the execution time for predicting both protein and promoter regions which is very promising. Table. 5 shows the number of datasets handled by AIS-INMACA. Figure. 6 and Figure. 7 shows the accuracy of prediction separately which is the important output of our work. Figure. 8 gives the prediction of extons

| Data Set | 54, Length DNA Sequence | 108, Length DNA Sequence | 162, Length DNA Sequence | 252, Length DNA Sequence | 354, Length DNA Sequence |
|---|---|---|---|---|---|
| Training Set Coding | 21.203 | 7,452 | 3,520 | 2,201 | 1,003 |
| Training Set Non Coding | 22.563 | 35,256 | 2,560 | 2,056 | 506 |
| Testing Set Coding | 22,569 | 21,023 | 15,564 | 3,002 | 689 |
| Testing Set Non Coding | 32,562 | 28,568 | 12,056 | 2,006 | 700 |

**Table 5:** Number of Data Sets Used

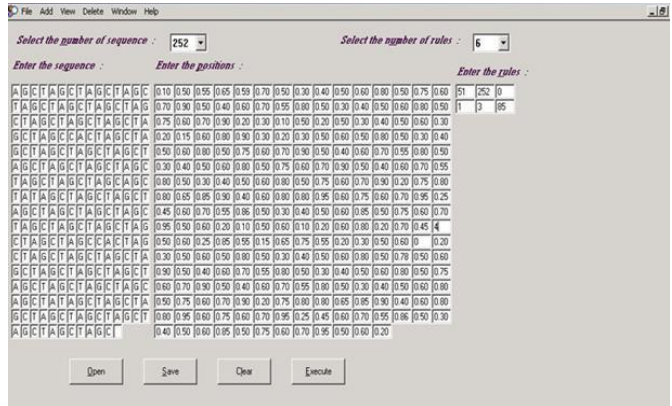

**Figure 4:** Training Interface

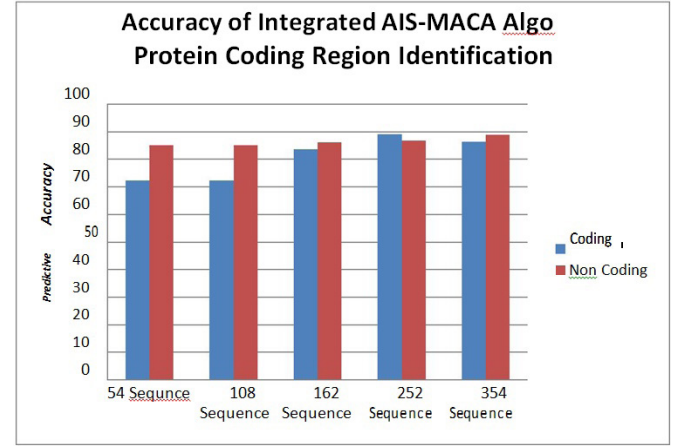

**Figure 6:** Predictive Accuracy for Protein Coding Regions

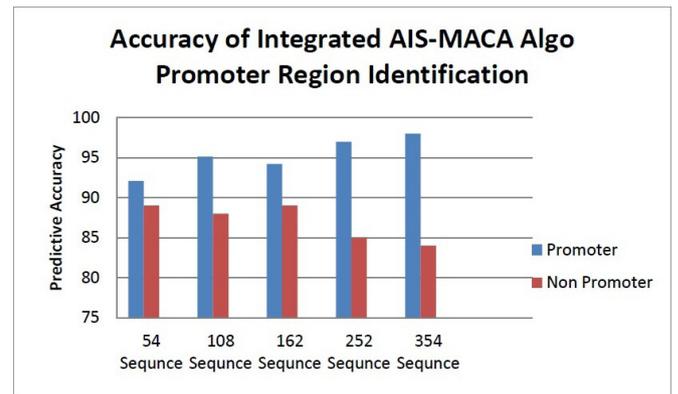

**Figure 7:** Predictive Accuracy for Promoter Regions

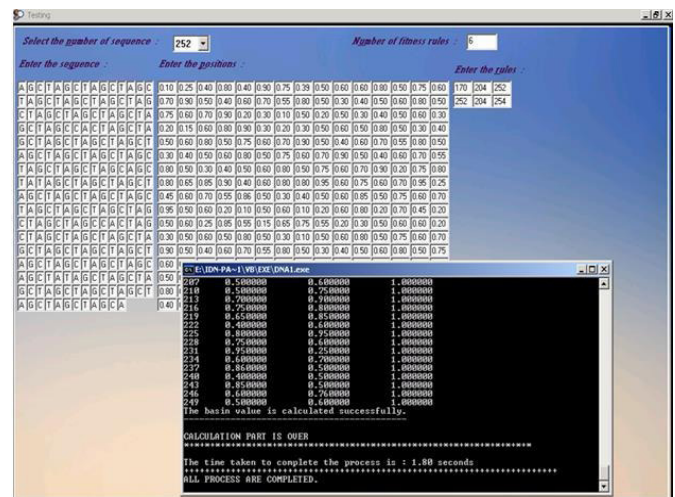

**Figure 5:** Testing Interface

| Size of Data Set | Prediction Time of Integrated Algorithm |
|---|---|
| 5000 | 1064 |
| 6000 | 1389 |
| 10000 | 2002 |
| 20000 | 2545 |

**Table 4:** Execution Time for prediction of both protein and promoter regions.

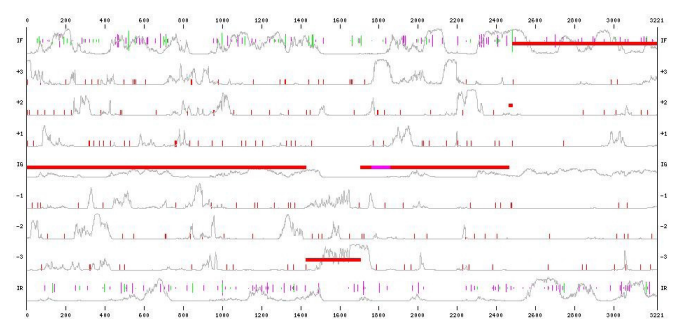

**Figure 8:** Exons Prediction

| Gene number | Element number | Exons/UTR | Strand | Left end | Right end | Length |
|---|---|---|---|---|---|---|
| 1 | 0 | Utr3 | + | 1 | 1402 | 1402 |
| 2 | 0 | Utr3 | - | 1404 | 1425 | 22 |
| 2 | 1 | Single | - | 1426 | 1704 | 279 |
| 2 | 0 | Utr5 | - | 1705 | 1760 | 56 |
| 3 | 0 | Utr5 | + | 1861 | 2464 | 604 |
| 3 | 1 | Initial | + | 2465 | 2482 | 18 |

**Figure 9:** Exons Boundaries Reporting

| Start | End | Score | Promoter Sequence |
|---|---|---|---|
| 268 | 313 | 0.99 | ATACTTTGGACAGAAAGAGGGTCCCAGTATAGCTTGGGCGACCTCTGAAA |
| 866 | 911 | 0.84 | CACCCTCTTGATTTTTTTTCCTTTCCAATTTATTCGGATCTGCTCTGGA |
| 1139 | 1184 | 0.80 | TGGGAATGACAGTCCCTAGCTTGGACCCTTAGACTGCCCGTTCTCATCCA |
| 1447 | 1492 | 0.86 | AGCCCTTGACAGGTCAGGTCCGCGGGACCCAGAATTCGCTTTGAGCCTAT |
| 1914 | 1959 | 0.91 | GGAGTTGCAGACTCAGGAATCAGGGAGCGGATACTGGGCGTGGACCCCGT |

**Figure 10:** Predictive Accuracy for Promoter Regions

from the given input graphically. Figure 9. gives the boundary of the location of extons. Figure 10. gives the promoter region prediction and boundary reporting.

## Conclusion

We have successfully developed a logical classifier designed with MACA and strengthened with AIS technique. The accuracy of the AIS-INMACA classifier is considerably more when compared with the existing algorithms which are 84% for protein coding and 90% for promoter region prediction. The proposed classifier can handle large data sets and sequences of various lengths. This is the first integrated algorithm to process DNA sequences of length 252,354. This novel classifier frame work can be used to address many problems in like protein structure prediction, RNA structure prediction, predicting the splicing pattern of any primary transcript and analysis of information content in DNA, RNA, protein sequences and structure and many more. We have successfully developed a frame work with this classifier which will lay future intuition towards application of CA in number of bioinformatics applications.